\documentclass[%
 reprint,
 amsmath,amssymb,
 aps,nofootinbib,
]{revtex4-1}
\usepackage{bm}
\PassOptionsToPackage{linktocpage}{hyperref}
\usepackage[hyperindex,breaklinks]{hyperref}
\usepackage{enumitem}
\usepackage{slashed}

\renewcommand{\theta}{\vartheta}
\newcommand{\p}{\partial}
\usepackage{array}
\usepackage{mathtools}
\DeclareMathOperator{\Tr}{Tr}
\usepackage{etoolbox}

\begin{document}

\title{Domestic Axion}

\author{Gia Dvali$^{1,2,3}$}
\author{Lena Funcke$^{1,2}$}
\email{Lena.Funcke@physik.uni-muenchen.de}
\affiliation{%
$^1$ Arnold Sommerfeld Center, Ludwig-Maximilians-Universit\"at, Theresienstra{\ss}e 37, 80333 M\"unchen, Germany
}%
 \affiliation{%
$^2$ Max-Planck-Institut f\"ur Physik, F\"ohringer Ring 6, 80805 M\"unchen, Germany
}%
 \affiliation{%
$^3$ Center for Cosmology and Particle Physics, Department of Physics, New York University, 4 Washington Place, New York, NY 10003, USA
}%

\date{\today}

\begin{abstract}
We attempt to identify a phenomenologically viable solution to the strong $CP$ problem in which the axion is composed entirely out of Standard Model fermion species. The axion consists predominantly of the $\eta'$ meson with a minuscule admixture of a pseudoscalar bilinear composite of neutrinos, $\eta_{\nu}$. The Peccei-Quinn symmetry is an axial symmetry that acts on the up quark and the neutrino species and is spontaneously broken by the QCD condensate of quarks as well as the condensate of neutrinos triggered by chiral gravitational anomaly. The up-quark mass is spontaneously generated by the neutrino condensate which plays the role of an additional composite Higgs doublet with the compositeness scale of the order of the neutrino masses. Such a scenario is highly economical: it solves the strong $CP$ problem, generates the up-quark and neutrino masses from fermion condensates and simultaneously protects the axion shift symmetry against gravitational anomaly. The phenomenology is different from the standard hidden axion case. One of the experimental signatures is the existence of a gravity-competing isotope-dependent attractive force among nucleons at (sub)micron distances.    
\end{abstract}


\maketitle

\section{Introduction}

The celebrated Peccei-Quinn (PQ) solution to the strong $CP$ problem \citep{Peccei1977} relies on the existence of a spontaneously broken chiral $U(1)_{\rm PQ}$ symmetry that is anomalous under the QCD gauge group. In such a case, the QCD $\theta$-term
\begin{equation}
\mathcal{L}_{\rm QCD}\supset\theta G\tilde{G}
\label{theta}
\end{equation} 
(where $G$ is the gluon field strength, $\tilde{G}$ is its dual, and irrelevant numerical coefficients are dropped as throughout the whole paper) becomes unphysical as it gets relaxed to zero by the pseudo-Goldstone boson of the $U(1)_{\rm PQ}$ symmetry, the axion $a_{\rm PQ}$ \citep{Weinberg1977-Wilczek1978}. The relaxation happens, because the $U(1)_{\rm PQ}$ symmetry, which acts on the axion as a shift symmetry,
\begin{equation}
 a_{\rm PQ} \rightarrow a_{\rm PQ} + {\rm const.}\, ,
\label{shiftaxion}
\end{equation}
is explicitly broken by the QCD anomaly. 

Within the Standard Model (SM), the simplest realization of such an anomalous chiral symmetry could have been achieved if one of the quark flavors, say the up quark, had no Yukawa coupling to the Higgs doublet.\footnote{One may argue that setting the Yukawa coupling constant of the up quark to zero creates another naturalness problem. This is a spurious argument, since setting a number protected by a symmetry to zero is not more unnatural than choosing it to be $\sim 10^{-5}$, putting aside that the gain of solving the strong $CP$ problem is enormous.} In such a case, the anomalous chiral symmetry would have been an axial $U(1)_{Au}$ symmetry acting on the up quark, 
\begin{equation}
u \rightarrow e^{i \alpha \gamma_5} u \,,   
\label{chiralu} 
\end{equation}  
where we combined the left-handed ($u_L$) and the right-handed ($u_R$) components of the up quark into a single Dirac fermion $u$. The corresponding current
\begin{equation}
j_{\mu}^{(u)}  \, = \, \bar{u}\gamma_{\mu}\gamma_5 u
\end{equation}
exhibits the anomalous Adler-Bell-Jackiw divergence with respect to QCD \citep{Adler1969-Bell1969},
\begin{equation}
\partial^{\mu} j_{\mu}^{(u)} \,  = \,   G\tilde{G}\,.    
\label{anomaly} 
\end{equation}  
Consequently, the vacuum $\theta$-angle can be removed by performing the chiral transformation (\ref{chiralu}) and becomes unphysical. 

Although sometimes this scenario is presented as being different from the PQ case, in reality it represents a particular version of the PQ solution \cite{Dvali2005_2}: the chiral symmetry is spontaneously broken by the QCD up-quark condensate and the role of the axion is played by the $\eta'$ meson. This degree of freedom describes the fluctuations of the phase of the quark condensate $\langle \bar{u}_L u_R \rangle \equiv  V^3e^{i \eta' / V}$, where $V^3$ is the vacuum expectation value (VEV) of its absolute value, which is set by the QCD scale, $|\langle \bar{u}_L u_R \rangle| = V^3  \sim \Lambda_{\rm QCD}^3$. In the absence of the up-quark Yukawa coupling constant (and when ignoring all other quark flavors), the $\eta'$ meson is getting its mass solely from the QCD anomaly \cite{'tHooft1976,Witten1979-Veneziano1979}.

Needless to say, such a solution to the strong $CP$ problem would be highly economical. Unfortunately, it does not work due to phenomenological reasons: the chiral perturbation theory indicates the need for an additional (on top of the QCD anomaly) contribution to the up-quark mass that breaks the chiral symmetry (\ref{chiralu}) (for a detailed discussion of this issue see \cite{Brambilla2014}). It is usually assumed that within the SM, the only possible source for the up-quark mass is the Yukawa coupling to the Higgs doublet, which of course is incompatible with the solution to the strong $CP$ problem as it breaks the chiral PQ symmetry (\ref{chiralu}) explicitly. 

By taking this standard road, one needs to implement the axion in form of a hypothetical degree of freedom from beyond the SM. This requires an introduction of a singlet scalar field with a very large VEV plus either a hypothetical heavy quark \cite{Kim1979-Shifman1979} or an additional Higgs doublet \cite{Zhitnitsky1980-Dine1981} (for a review see, e.g., \cite{Peccei2006}). 

We shall not take this road. Instead, in the present paper we shall ask whether the needed contribution to the up-quark mass can be provided by a neutrino condensate in such a way that it could break the chiral PQ symmetry (\ref{chiralu}) {\it spontaneously}. The existence of such a neutrino condensate -- as we shall explain -- follows from a very general assumption about the topological structure of the vacuum due to gravitational chiral anomaly.

In order to better capture its novel aspects, it is useful to confront the present scenario with the original Weinberg-Wilczek axion case \cite{Weinberg1977-Wilczek1978}. For the existence of chiral PQ symmetry, it is a necessary condition that different quarks get their masses from different Higgs doublets. In the original axion scenario, this is accomplished by coupling the up and down quarks to two distinct Higgs doublets $H$ and $H'$,
\begin{equation} 
\mathcal{L}_{\rm PQ} \, = \,  H' \bar{Q}_L u_R \, + \, H \bar{Q}_L d_R \, + ...\, ,
\label{twoHiggses1}
\end{equation}
where $Q_L \equiv (u_L,d_L)$ is the doublet of left-handed quarks. This decoupling of some quarks from a particular Higgs doublet is justified by the chiral PQ symmetry, $H \rightarrow e^{i\alpha} H,~ H' \rightarrow e^{i\alpha} H',~ (\bar{Q}_L u_R)  \rightarrow e^{- i\alpha} (\bar{Q}_L u_R), ~ (\bar{Q}_L d_R)  \rightarrow e^{- i\alpha} (\bar{Q}_L d_R)$. In this scenario, the axion comes predominantly from the phase of the neutral Higgs with a smaller VEV, but since the VEV is around the weak scale, such an axion is ruled out experimentally. 

In our model, it remains true that the different quarks are getting masses from different Higgs doublets, but the additional doublet is provided by the SM itself: it is a neutrino condensate. The SM fermion composition of the emerging PQ axion is the reason why we will call this axion a \textit{domestic axion}.
 
The simplest prototype effective Lagrangian describing the Domestic Axion idea is
\begin{equation} 
\mathcal{L}_{\rm DA} \, = \, \frac{f}{\Lambda_G^2} (\bar{\nu}_R L) \bar{Q}_L u_R \, + \, H \bar{Q}_L d_R \,  + ...\, , 
\label{twoHiggses2}
\end{equation}
where $L\equiv (\nu_L, e_R)$  is the lepton doublet and the scale $\Lambda_G$ and the invariant function $f$ are provided by gravity and will be discussed below.  
 
Thus, the additional doublet $H'$ of the original PQ model is replaced by an effective doublet composed out of the lepton doublet and the right-handed neutrino, $H' \rightarrow (\bar{\nu}_R L)$. In this minimal realization,  the PQ symmetry is the chiral symmetry acting both on quarks as well as on neutrinos, $(\bar{\nu}_R L)  \rightarrow e^{i\alpha} (\bar{\nu}_R L)$, and is spontaneously broken by both condensates.  

The crucial ingredient here is the condensate of the composite doublet $\langle \bar{L} \nu_R\rangle  =  \langle \bar{\nu}_L \nu_R\rangle  \neq 0$, imposed by the gravitational chiral anomaly. The role of this condensate is to spontaneously generate the mass of the up quark, but the contribution from its phase, called the $\eta_{\nu}$ boson, to the axion is negligible. Instead, the axion is coming almost entirely from the $\eta'$ meson of QCD, because the breaking of chiral PQ symmetry is predominantly accomplished by the QCD condensate of quarks, which is much larger than the gravitationally induced neutrino condensate. 

The roles of the pseudo-Goldstone bosons are split in the following way: the $\eta'$ meson is getting its mass from the QCD anomaly and becomes an axion, whereas the $\eta_{\nu}$ boson is getting its mass from the gravitational anomaly and ``sacrifices" itself for protecting the shift symmetry of the $\eta'$ meson against the gravitational anomaly via the mechanism of \cite{Dvali2005_2,Dvali2013}. The crucial point that makes our neutrino-composite doublet compatible with experimental bounds is that it is very ``fat": its extremely low compositeness scale makes it to contribute only into very soft processes and to efficiently decouple from the hard high-energy processes.\\

Such a Dometic Axion scenario has the following obvious advantages:
   
\begin{enumerate}[label={(\arabic*)}]
\item It provides the axion without any need of postulating the existence of new hypothetical species. 
\item The axion is automatically immune to the gravitational anomaly and its shift symmetry (\ref{shiftaxion}) is broken exclusively by the QCD effects \cite{Dvali2005_2, Dvali2013}. 
\item The neutrino condensate that breaks PQ symmetry is also the source of the neutrino masses, via the mechanism of \cite{Dvali2016}. 
\end{enumerate}
 
Thus, the present scenario connects the solution to the strong $CP$ problem to the origin of the neutrino masses, without the need for new species, and simultaneously protects the axion solution against gravity.

\section{Neutrino protection against gravitational anomaly}  

Before presenting the complete model, let us briefly elaborate on each of the above topics and review the previous results that we shall use.\\
    
First, let us discuss the gravitational threat. In order for the axion to relax the $\theta$-term to zero, the axion shift symmetry (\ref{shiftaxion}) must be broken exclusively by the QCD effects via the anomaly. However, there is an old believe that quantum gravity effects can generate an additional breaking of the axion shift symmetry (\ref{shiftaxion}) and therefore ruin the axion solution to the strong $CP$ problem (see, e.g., \cite{Kamionkowaski1992}). 

The necessary and sufficient conditions for the possibility of such an explicit breaking were identified in \cite{Dvali2005_2}, where -- by reformulating the axion solution in the language of a three-form Higgs effect -- the breaking of the axion shift symmetry by gravity was linked to the gravitational chiral anomaly and to the gravitational topological susceptibility of the vacuum.\footnote{As shown in \cite{Dvali2005}, the direct connection between the topological susceptibility and the generation of the mass gap in the anomalous current is a very general phenomenon and goes well beyond gravity.} Namely, the condition is that -- in a theory without an axion -- the gravitational topological vacuum susceptibility in the limit of vanishing momentum $q$ is nonzero, 
\begin{align}\label{RRcorr}
\langle R\tilde{R},R\tilde{R}\rangle_{q\to 0} &\equiv\lim\limits_{q \to 0}\int d^4 x e^{iqx} \langle T[R\tilde{R}(x)R\tilde{R}(0)]\rangle\nonumber\\&= \mathrm{const}\neq 0\, ,   
\end{align}
where $R$ is the Riemann tensor and $\tilde{R}$ is its dual. Note that this condition is equivalent to the statement that the gravitational analog of the $\theta$-term, 
\begin{equation}
\mathcal{L}_G\supset\theta_G R\tilde{R} \, ,
\label{gravitheta} 
\end{equation}
is physical. 

The existence of a nonvanishing topological vacuum susceptibility in pure gravity is currently an open question. If it is zero, then the gravitational anomaly poses no danger to the axionic shift symmetry \cite{Dvali2005_2}. But, if it is nonzero, one has to face the consequences. What we want to show is that in such a case the gravitational danger comes with a built-in protection mechanism, which does not only eliminate itself, but as a bonus identifies the viable axion candidate within the SM in form of the $\eta'$ meson. \\ 

Thus, we shall assume that the above condition -- i.e., in the absence of an anomalous current, gravity gives rise to (\ref{RRcorr}) -- is fulfilled and consequently the threat to the axion solution to the strong $CP$ problem from gravity is real. This introduces a new gravitational scale in the problem, $\Lambda_G$, which sets the scale of the correlator (\ref{RRcorr}). At the level of our discussion, $\Lambda_G$ is a free parameter, solely constrained by phenomenological requirements. One thing that we can expect about this scale is that it must be strongly suppressed with respect to the Planck scale. This is normal for the infrared (IR) scales generated by nonperturbative effects, such as instantons or virtual black holes. However, in this paper we shall not commit to any particular microscopic origin of the correlator (\ref{RRcorr}), which can easily be {\it intrinsically quantum gravitational} rather than semi-classical. Another natural assumption is that the effective low-energy interactions generated by this IR physics must become irrelevant in short-distance processes at energies $E \gg \Lambda_G$, i.e., their contribution must sharply diminish for $\Lambda_G /E \ll 1$.\footnote{Later, for the phenomenological estimates we shall parameterize our ignorance about the high-energy behavior of the effective interactions generated by nonperturbative gravitational physics by a power-law dependence on ${\Lambda_G / E}$.}

In such a scenario, as explained in \cite{Dvali2005_2}, in the absence of an axion (or a massless fermion), there exist two physically observable theta parameters, one from QCD (\ref{theta}) and one from gravity (\ref{gravitheta}). Consequently, after the axion is introduced, it can only cancel a single combination of the two $\theta$-terms, whereas the other combination remains physically observable. Hence, the strong $CP$ problem is not solved.  

In this situation, as a possible protection mechanism, it was suggested in \cite{Dvali2005_2} to take into account some fermions (e.g., neutrinos) with zero bare mass. In such a case, there always exists a chiral symmetry, which is anomalous with respect to gravity. For example, for a single Dirac neutrino flavor, we have an axial $U(1)_{A\nu}$ symmetry,
\begin{equation}
\nu \rightarrow e^{i \alpha \gamma_5} \nu \, ,
\label{chiralnu} 
\end{equation} 
with the corresponding axial current
\begin{equation}
j_{\mu}^{(\nu)} = \bar{\nu} \gamma_{\mu}\gamma_5\nu\, .
\label{currentnu} 
\end{equation}
Due to chiral gravitational anomaly \cite{Delbourgo1972-Eguchi1976-Deser1980-AlvarezGaume1984}, the current has an anomalous divergence,
\begin{equation}
\partial^{\mu} j_{\mu}^{(\nu)} \,  = \,   R\tilde{R}   
\label{anomalynu} 
\end{equation} 
and -- just like in QCD with a massless quark -- the gravitational $\theta$-term (\ref{gravitheta}) can be eliminated by an axial transformation of the neutrino (\ref{chiralnu}). As a result, the gravitational topological susceptibility (\ref{RRcorr}) vanishes and gravity generates a mass gap in the neutrino sector, so that the axion potential is not affected. This mechanism was implemented in details as the axion protection mechanism against gravity in \cite{Dvali2013}. One of the predictions of this scenario is the existence of a pseudo-Goldstone boson, $\eta_{\nu}$, which corresponds to the neutrino axial current \eqref{currentnu}. The $\eta_{\nu}$ boson represents a collective excitation of the neutrino condensate phase and plays the role closely analogous to the $\eta'$ meson of QCD, which gets its mass from the QCD anomaly \eqref{anomaly}. 

The next step was undertaken in \cite{Dvali2016}, where it was suggested to identify the neutrino condensate -- triggered by gravitational anomaly -- as the unique source of all the neutrino masses. This fixes the scale of the condensate in the $\sim 0.1$~eV range.\\

In all the above studies, it was assumed that the axion that solves the strong $CP$ problem is coming from some unspecified beyond the SM sector. In the present paper, we would like to suggest a much more economical possibility: we would like to propose the scenario in which the neutrino condensate also generates the mass of the up quark spontaneously.

In such a scenario, the attractive feature is that the role of the PQ symmetry is played by a combination of the axial symmetries (\ref{chiralu}) and (\ref{chiralnu}) acting on the up quark and on the neutrinos, respectively. This symmetry is free of gravitational anomaly and is anomalous solely with respect to QCD. It is spontaneously broken by the QCD up-quark condensate as well as by the condensate of the neutrinos. Since the quark condensate is dominant, the corresponding axion mostly consists of the QCD $\eta'$ meson with a small admixture of $\eta_{\nu}$. The orthogonal combination, which consist mostly of $\eta_{\nu}$ with a small admixture of $\eta'$, gets its mass from the gravitational anomaly.

\section{The Model}

\subsection{Anomalous $U(1)_G$ and $U(1)_{\rm PQ}$ symmetries}

Let us now describe our model in more details. The key postulate is that the masses of some quarks are generated by their couplings to the neutrino condensate as opposed to the Yukawa couplings to the SM Higgs. The neutrino condensate acts as an additional composite Higgs doublet, and this allows the Lagrangian to be invariant under a chiral PQ symmetry that is anomalous with respect to QCD. For solving the strong $CP$ problem in this way, it is unimportant which quarks are getting their masses from the neutrino condensate, but it would be natural to employ the light quarks. 

We shall start with a  minimal scheme in which only the up quark and a single neutrino flavor are involved. We thus set to zero the Yukawa coupling constants of the Higgs doublet to the up quark and to one of the three neutrino flavors. We shall assume that the masses of all the other fermions are generated in a standard way through their Yukawa couplings to the Higgs VEV and we shall exclude them from our considerations. These additional fermions can be easily integrated back in without affecting the essence of the Domestic Axion scenario, and we will discuss this possibility later.  

The decoupling of the up quark and the neutrino from the Higgs doublet creates new global symmetries. At the perturbative level, gravity treats all the fermion species democratically; thus, it effectively sees the three colors of the left- and right-handed up-quark pairs and one left- and right-handed neutrino pair forming a representation of  the $U(4)_L\times U(4)_R$ flavor symmetry group.\footnote{Note, since all the fermions can be written in the left-handed basis (with right-handed fermions $\psi_R$ replaced by left-handed anti-fermions $\psi_{cL}$), they can be viewed as a fundamental representation of the $U(8)$ flavor group. However, the Lorentz and gauge invariant bilinear order parameters of the type $\bar{\psi}_L \psi_R$ form the bifundamental representations of the $U(4)_L\times U(4)_R$ group.}

Before taking into account the quantum anomalies, the QCD and electromagnetic gauge interactions break this symmetry explicitly down to the following subgroup:
\begin{align}
 {\mathcal  G} \equiv  &\; SU(3)_{\rm color}\times U(1)_{\rm EM} \times & \nonumber \\
  &\;\times U(1)_{Vu} \times U(1)_{V\nu} \times U(1)_{Au} \times U(1)_{A\nu} \, ,
\label{subgroup} 
\end{align}
where $SU(3)_{\rm color}$ is a color group and $U(1)_{V u}$ and $U(1)_{V\nu}$ are the vector-like quark (baryon) and neutrino (lepton) number symmetries, respectively. Since we have ignored other fermion species, the electromagnetic symmetry $U(1)_{\rm EM}$ acts essentially as the gauged version of the up-quark number symmetry $U(1)_{Vu}$.  

The asymmetry between the left- and right-handed fermion species in the SM is only created by the weak gauge interaction. Since we are interested in very low energy phenomena, we shall ignore the effects that break the left-right symmetry. 
 
Finally, $U(1)_{Au}$ and $U(1)_{A\nu}$ are the quark and neutrino axial symmetries given by (\ref{chiralu}) and (\ref{chiralnu}), respectively. The following combination of these symmetries,
\begin{equation} 
u  \rightarrow e^{i \alpha \gamma_5} u \, ,  ~~  \nu \rightarrow e^{i \alpha \gamma_5} \nu \,,
\label{gravichiral} 
\end{equation}
is anomalous with respect to gravity, and we shall denote it by $U(1)_{G}$. The corresponding current
\begin{equation}
j_{\mu}^{(G)} \, = \, \sum_a \bar{u}^a\gamma_{\mu} \gamma_5 u_a  \, + \,   \bar{\nu}\gamma_{\mu} \gamma_5 \nu\, ,
\end{equation}
exhibits the anomalous divergence \eqref{anomalynu}
\begin{equation}
\partial^{\mu} j_{\mu}^{(G)} \,  = \,   R\tilde{R}\,.    
\label{anomalynuq} 
\end{equation}  
Note that the anomalous $U(1)_G$ symmetry contains an anomaly-free $Z_{8}$ subgroup corresponding to the discrete values of the phase parameter $\alpha \,  = \, {\pi \over 4}n$ with $n$ being an arbitrary integer.
   
Another important symmetry is the orthogonal combination of $U(1)_{Au}$ and $U(1)_{A\nu}$,  
\begin{equation} 
u_a \rightarrow e^{i \alpha \gamma_5} u_a \, ,  ~~  \nu  \rightarrow e^{- i 3\alpha \gamma_5} \nu \, , 
\label{graviPQ} 
\end{equation}  
which we shall denote by $U(1)_{\rm PQ}$. This symmetry is free of gravitational anomaly, but it is anomalous with respect to QCD, and the corresponding current
\begin{equation}
j_{\mu}^{\rm (PQ)} \, = \,\sum_a \bar{u}^a\gamma_{\mu} \gamma_5 u_a  \, - \, 3\, \bar{\nu}\gamma_{\mu} \gamma_5 \nu
\end{equation}
exhibits the anomalous divergence \eqref{anomaly}
\begin{equation}
\partial^{\mu} j_{\mu}^{\rm (PQ)} \,  = \,   G\tilde{G}\,.    
\label{anomaly2} 
\end{equation}  
Thus, this symmetry is the right candidate for the PQ symmetry. Note, although both symmetries (\ref{gravichiral}) and (\ref{graviPQ}) include a $U(1)_{Au}$ component and therefore are anomalous with respect to QCD, we identify (\ref{graviPQ}) as the PQ symmetry since it is the one that is anomaly-free with respect to gravity.

\subsection{Generation of up quark mass}   

Let us now discuss the effective interaction that is induced by the gravitational anomaly and is responsible for generating the mass gaps for the Goldstone bosons as well as for the fermions. Since we use the anomaly-free symmetries as the guideline, we shall consider interactions that are invariant with respect to (\ref{subgroup}). 

The pattern of chiral symmetry breaking is determined by minimization of an effective potential for the quark and the neutrino order parameters, $X_u \equiv (\bar{u}_L u_R)$ and $X_{\nu} \equiv (\bar{\nu_L}\nu_R)$. This effective potential consists of the ordinary QCD part and the part generated by gravity. The QCD part consists of the effective potential that induces the quark condensate and breaks the axial $U(1)_{Au}$ symmetry (\ref{chiralu}) spontaneously as well as the 't Hooft-type interactions that break this symmetry explicitly and contribute into the mass for the $\eta'$ meson. Likewise, the effective potential induced by gravity can be split into the part that breaks $U(1)_G$ symmetry spontaneously and the one that breaks it explicitly. 
 
The parts that are responsible for spontaneous breaking are given by some unknown polynomial consisting of (generically infinite series of) phase-independent invariants, such as, $X_u^+X_u$ and $X_{\nu}^+X_{\nu}$. Its explicit form is unimportant for our purposes. It suffices to know that the minimum of this effective potential is achieved for a nonvanishing VEV of the neutrino condensate, $\langle X_{\nu} \rangle = v^3 e^{i{\langle \eta_{\nu} \rangle / v}}$, where $v$ is the characteristic scale of the condensate and $\langle \eta_{\nu} \rangle $ is the VEV of its phase. The scale $v$ is {\it a priori} unknown, and we must treat it as a free parameter. We do not expect it to be very far from the scale $\Lambda_G$ that sets the scale of the correlator (\ref{RRcorr}), although it can be parametrically different. So we shall assume $v \sim \Lambda_G$. 
  
Gravity is expected to also trigger a condensate for $X_{u}$ of similar order of magnitude, but this is just a tiny correction to the condensate of $X_{u}$ triggered by QCD, $\langle X_{u} \rangle = V^3 e^{i{\langle \eta' \rangle / V}}$, where $V$ is of order the QCD scale.  
  
If the considered effective potential consisted solely of the spontaneous-breaking part, the phase degrees of freedom, $\eta_{\nu}$ and $\eta'$, would be {\it exactly  massless} Goldstone bosons. However, from the anomaly and topology we know that the mass gaps in both of these Goldstones must be generated. In particular, QCD generates a mass gap in $\eta'$.  At the level of the effective potential, this can be modeled by a 't Hooft-type vertex, which for a single quark case is just a linear term in $X_{u}$ multiplied by an arbitrary function of the phase-independent invariant $X_u^+X_u$.\footnote{Note that we do not commit here to the assumption that the main source of the $\eta'$ mass in QCD are instantons. As it is well known \cite{Witten1979-Veneziano1979}, for a large number of colors the Witten-Veneziano mechanism is expected to give the dominant contribution.  For us, the 't Hooft like structure -- regardless of its underlying origin -- is a useful parameterization of the  symmetry properties of the effective vertex that explicitly breaks the anomalous $U(1)_{Au}$ chiral symmetry to an anomaly-free discrete subgroup and generates the pseudo-Goldstone mass. In case of a single quark flavor, the anomaly-free symmetry is $Z_2$ and this uniquely fixes the structure of the vertex in form of a linear term in $X_u$ times an arbitrary function of phase-independent invariants.}

Likewise, gravity generates the mass gap for a particular superposition of Goldstones corresponding to the $U(1)_G$ symmetry. Correspondingly, the effective Lagrangian generated by gravity on top of the standard QCD effects must contain additional interaction terms among $X_u$ and $X_{\nu}$, which break the anomalous $U(1)_G$ symmetry explicitly and generate the pseudo-Goldstone masses. The same interaction shall contribute into the spontaneous generation of the masses of the up quark and the neutrino. Let us show how this happens.

In order to be concrete, we model the gravity-induced interaction by the following vertex: 
\begin{align} 
\mathcal{L} _{G} \, = \,   { 1 \over \Lambda_G^2} (X_u X_{\nu}) \, 
f(X_u^+X_u, X_{\nu}^+X_{\nu},...) \, + \, {\rm h.c.},  
\label{vertex}
\end{align}  
where $\Lambda_G$ is the IR scale of gravity.\footnote{For example, the above interaction can be thought as being generated from a $SU(8)$-invariant 't Hooft-type gravitational vertex,
\begin{align} 
{ 1 \over \Lambda_G^8} &(\bar{u}^1_L u_{1R}) 
(\bar{u}^2_L u_{2R}) (\bar{u}^3_L u_{3R}) 
(\bar{\nu}_{L}\nu_R)  \, + \, {\rm h.c.},  
\label{vertex8}
\end{align} 
which explicitly breaks (\ref{gravichiral}) but respects (\ref{graviPQ}), after dressing it by QCD effects, which alone would break $U(1)_{Au}$ but not $U(1)_{A\nu}$. The combined effects of generated terms do not leave any unbroken continuous chiral symmetry. 

We need to stress again that we are bringing the analogy with the 't Hooft vertex exclusively because of the $Z_8$-symmetry structure of the vertex. We are {\it not} assuming that the gravitational vertex is necessarily generated from semi-classical physics, such as gravitational instantons. Rather, its origin can be deeply quantum gravitational.}  

The function $f(X_u^+X_u, X_{\nu}^+X_{\nu},...)$ is some unknown dimensionless function of the phase-independent quark and neutrino invariants, $X_u^+X_u$ and $X_{\nu}^+X_{\nu}$, scaled by the parameter $\Lambda_G$ (for more discussion about $f$ see App. \ref{App:Condensate}). For the phenomenological consistency of our scenario, we need to impose the following constraint on this function:
\begin{align} 
\langle X_u \rangle \biggl\langle {\partial^2  \mathcal{L} _{G}  \over \partial \bar{u}_L\partial u_R} \biggr\rangle \, &=  \, \xi \,\langle \mathcal{L} _{G} \rangle \sim \nonumber\\ &\sim \, \xi \langle X_{\nu} \rangle \biggl\langle {\partial^2  \mathcal{L} _{G}  \over \partial \bar{\nu}_L\partial \nu_R} \biggr\rangle \, ,
\label{conditionX}
\end{align} 
with $\xi \sim 10^{7}$. This means, the VEV of the derivatives of the function $f$ with respect to $X_u$ must be much larger than the other expectation values, i.e., the function $f$ must be steep (or highly curved) in $X_u$ direction. As we shall see in a moment, this condition replaces the fine tuning of the up-quark Yukawa coupling constant in the standard scenario. In the standard case, the tuning of the Yukawa coupling constant sets the hierarchy between the up-quark mass and the Higgs VEV, whereas in our case, $\xi$ sets the hierarchy between the up-quark and the neutrino masses. As it will become clear later, the same condition also guarantees that the Goldstone bosons are not entering the strong coupling regime. \\

Since the neutrino condensate contributes into the spontaneous breaking of the $U(1)_{\rm PQ}$ symmetry, it provides an additional non-QCD contribution to the up-quark mass through the vertex (\ref{vertex}). An effective up-quark mass term is obtained by replacing all fermion bilinears by their VEVs while keeping the two quark legs free. We get
\begin{equation} 
\biggl\langle {\partial^2  \mathcal{L} _{G}  \over \partial \bar{u}_L\partial u_R} \biggr\rangle\, \bar{u}_L u_R  \simeq  \, \xi \, { v^{3} \over \Lambda_G^2} \, \langle f \rangle \, \bar{u}_L u_R  \,  + \, ...  \, ,
\label{massupquark}
\end{equation}   
where we took into account the condition (\ref{conditionX}). The resulting up-quark mass thus is
\begin{equation}
m_u \simeq \,  \xi \, v \left ({v \over \Lambda_G}\right)^2 \langle f \rangle.
\end{equation}
By the same estimate, the contribution to the neutrino masses from the above vertex is 
\begin{equation}
m_{\nu} \simeq  \,  v \left ({v \over \Lambda_G}\right)^2 \langle f \rangle.
\label{massneutrino}
\end{equation} 
Notice, despite the fact that the QCD-induced up-quark condensate is large, when inserted into the gravitational vertex it must be effectively cut-off around the scale $v$, since this is the scale of the softness of the effective vertex (\ref{vertex}). Thus, we have effectively represented all the condensates by the scale $v \sim \Lambda_G$. Then, the hierarchy between the neutrino and the up-quark masses is controlled by the parameter $\xi$. For a phenomenologically acceptable value of the up-quark mass we need to choose $\xi \sim 10^7$ for an up-quark mass of $m_u\sim$~MeV and neutrino masses of about $m_\nu \sim 0.1$~eV.\\

Although the above choice of the parameter $\xi$ may seem somewhat unnatural, it is much milder than the fine tuning of parameters required for achieving a more modest goal in the standard hidden axion models. In these models, first, one anyway needs to fine-tune the Yukawa coupling constant of the up quark to the value $\sim 10^{-5}$, and second, given the phenomenological lower bound on the PQ scale, $\sim 10^{9}$~GeV \cite{Raffelt2008}, one has to fine-tune the mass-square term of the Higgs boson relative to the PQ scale by a factor of $\sim 10^{-14}$. This putting aside that these standard scenarios have nothing to say either about the protection of the axion solution against gravity or about the origin of the neutrino masses. So, in this light and given the goals we aim to achieve, the relatively large choice of $\xi$ may not be such a big price to pay after all.\footnote{Also when comparing to the fine tuning of the up-quark Yukawa constant in the standard scenario, it is important to stress that $\xi$ is a coefficient of a very high dimensional operator and its tuning amounts to much milder tuning when translated in terms of mass scales, because of high-power sensitivity.}

\subsection{Emergence of domestic axion and graviaxion}

The vertex (\ref{vertex}) in combination with the standard QCD contribution explicitly breaks all continuous chiral symmetries. Correspondingly, both would-be Goldstone bosons, $\eta'$ and $\eta_{\nu}$, become massive pseudo-Goldstones. 

In order to evaluate their masses, we shall replace the absolute values of the fermion bilinears by their VEVs and express their phases though the corresponding pseudo-Goldstone modes. We thus write $\langle \bar{u}_L u_R \rangle =  V^3e^{i {\eta' / V}}$ and $\ \langle \bar{\nu}_{L}\nu_{R} \rangle \,  = \, v^3e^{i {\eta_{\nu} / v}}$, where $V$ and $v$ are the scales of the two condensates introduced above.

As already mentioned, the neutrino condensate forms due to nonperturbative gravitational effects, whereas the up-quark condensate is dominantly triggered by QCD effects with a negligible gravitational contribution. Thus, the scale $V$ is given by the QCD scale, $V = \Lambda_{\rm QCD}$. Nevertheless, when inserted into the vertex (\ref{vertex}), we have to set the VEV of the absolute value also for $X_u$ to be of order $v^3$. The reason is the same UV softness of the vertex (\ref{vertex}) as explained earlier. When we terminate the external legs of the vertex into the VEVs, we should keep in mind that the contribution freezes out above a certain critical value of the VEV. This value corresponds to the momentum above which the vertex (\ref{vertex}) is resolved and ``melts". We assumed this scale to be given by $v \sim \Lambda_G$. Hence, when we plug the quark condensate into the vertex, we must effectively replace it with $\langle \bar{u}_L u_R \rangle =  v^3e^{i {\eta' / V}}$. Notice, the decay constant of the $\eta'$ is still given by $V$, because this is just an information about the canonical normalization of the pseudo-Goldstone mode. 

Plugging the above expressions for the fermion bilinears into (\ref{vertex}), we obtain
\begin{align} 
\mathcal{L}_{G} \,&= \,   { v^{6} \over \Lambda_G^2} \langle f \rangle \; {\rm cos} \left({\eta' \over V} + {\eta_{\nu} \over v} \right ).
\label{massgraviaxion1}
\end{align} 
Expanding the cosine yields an effective mass term,
\begin{align} 
\mathcal{L}_{\rm mass} \,&= - \, {1 \over 2} m^2_{G}  \,  a_{G}^2\, ,
\label{massgraviaxion2}
\end{align} 
for one combination of the Goldstone modes,
\begin{equation}
a_G \equiv  {\eta_{\nu} + \eta' \epsilon \over \sqrt{1 + \epsilon^2}}\, ,
\label{graviaxion} 
\end{equation} 
with the mass
\begin{equation}
m_{G}^2 \, = \, {v^{4}  \over \Lambda_G^2} \:(1 + \varepsilon^2) \langle f \rangle \simeq 
 v^2 \left ({v \over \Lambda_G}\right)^2\langle f \rangle\;.
\label{massgraviaxion3}
\end{equation}
Here we have taken into account the smallness of the parameter $\varepsilon \equiv  v/V$. Since the function $f$ only depends on real invariants, it does not break any of the $U(1)$ symmetries and contributes to the Goldstone potential only in form of an overall factor.  
 
From (\ref{massgraviaxion2}) it is clear that the mode $a_G$ is the pseudo-Goldstone boson that gets its mass from the gravitational anomaly and screens the gravitational $\theta$-term. It consists mostly of the neutrino-composite pseudoscalar $\eta_{\nu}$ with a small ($\sim \epsilon$) admixture from the $\eta'$ meson of QCD. In the absence of the QCD anomaly, the above mode would be a true mass eigenstate, while the orthogonal combination, $ {(\eta'  - \eta_{\nu} \epsilon) / \sqrt{1 + \epsilon^2}}$, that is a Goldstone boson of the spontaneously broken $U(1)_{\rm PQ}$ symmetry, would remain exactly massless. 
 
However, this is not the case. The $U(1)_{\rm PQ}$ symmetry is anomalous with respect to QCD and the Goldstone bosons are also getting a mass from this anomaly. Note that it is only the $\eta'$ component that couples to QCD and contributes into the QCD chiral anomaly. This component is getting a mass through the QCD mechanism and solves the strong $CP$ problem.\footnote{Since $a_{\rm PQ}$ is mostly $\eta'$, its mass generation can be described in terms of the Witten-Veneziano mechanism \cite{Witten1979-Veneziano1979}. The only difference is that, since in our case the up-quark mass is generated spontaneously, $a_{\rm PQ}$ screens the $\theta$-term entirely and solves the strong $CP$ problem.}

As a result, the mass matrix takes the form (see App. \ref{App:MassMatrix} for an alternative derivation)
\begin{equation}
\mathcal{L}_{\rm mass} \, = \,- \, {1 \over 2} m_{\eta'}^2  \eta'^{2} \,  - \, {1 \over 2} m^2_{G} (\eta_{\nu} \, + \, \epsilon \eta')^2 \, .
\label{masses1}
\end{equation} 
As we can see, the mixing angle between $\eta'$ and $\eta_{\nu}$ is absolutely minuscule ($\sim \epsilon \,m_{G}^2/ m_{\eta'}^2 \sim  \epsilon^3$). Correspondingly, up to the mixing of order $\epsilon^3$, they are the true mass eigenstates, 
\begin{equation}
a_{\rm PQ} \, = \, \eta'  + {\mathcal O}(\epsilon^3) \eta_{\nu} \, ,~~~~
a_{G} \, = \, \eta_{\nu}\,  + \, {\mathcal O}(\epsilon^3) \eta' \,,
\label{states1}
\end{equation}
with the masses equal to $m_{\eta'}$ and $m_{G}$, respectively.

The boson $a_{\rm PQ}$ represents a perfect domestic PQ axion, but with one advantage: unlike ordinary PQ symmetries, in our case the $U(1)_{\rm PQ}$ symmetry, which shifts the axion (\ref{shiftaxion}), is free of gravitational anomaly and hence is protected against gravitational destabilization \cite{Dvali2005_2,Dvali2013}. This is also demonstrated in App. \ref{App:MassMatrix} in the language of a gauge three-form \cite{Dvali2005_2}.

The $a_{G}$ boson, which is mostly composed of $\eta_{\nu}$ and gets its mass from the gravitational anomaly, will be referred to hereafter as a {\it graviaxion}. For $\langle f \rangle \sim 1$, the mass \eqref{massgraviaxion3} of the graviaxion is of the same order as the neutrino mass and is given by the scale $\Lambda_G$. Notice, since the function $f$ is independent of phases, it only contributes into the Goldstone masses as an overall factor.  Therefore, in order to create a hierarchy between the neutrino and the up-quark masses -- without simultaneously pushing $m_{G}$ above the scale $v$ -- we need to take a large $\xi$ while keeping the VEV of $f$ to be of order one.

\section{Phenomenology} 

\subsection{Axion is $\eta'$}

Since in our model the axion $a_{\rm PQ}$ is predominantly $\eta'$, its phenomenological implications are essentially the same as the ones of the usual $\eta'$ meson, with the only difference of a tiny admixture of the $\eta_{\nu}$ boson. Correspondingly, the phenomenology of the axion will essentially be similar to the phenomenology of the standard $\eta'$. Most importantly, this implies that axion search experiments will not find any axion particle, but the axion was already discovered in 1964 \cite{Kalbfleisch1964-Goldberg1964}. All experiments designated for the discovery of hidden axions \cite{Graham2015} can only discover axion-like particles (ALPs) but not the axion itself. In our case, such an ALP is the graviaxion.

\subsection{Phenomenology of graviaxion}

Our other predicted pseudo-Goldstone boson, the graviaxion $a_{G}$, predominantly consists of $\eta_{\nu}$ and gets its mass from the gravitational anomaly. In this way, this boson protects the QCD $\eta'$-axion from gravity. The phenomenology of $\eta_\nu$ and of the entire modified neutrino sector from the previous paper \cite{Dvali2016} will essentially carry over to the present model, and we shall not repeat it here. Most of the new phenomena will manifest themselves in various very soft processes of extremely low momentum exchange.

One additional new low-energy process not mentioned in \cite{Dvali2016} is the graviaxion-to-photon conversion. Since the direct contribution from $\eta'$ into the graviaxion is very strongly suppressed, the dominant communication of the graviaxion to the photon is through the virtual charged particles (quarks and leptons) to which the graviaxion couples through the soft gravitational vertex. If we assume a maximally generic form of such a vertex, the least suppression factor we get can be estimated to be $(v/m_e)^3$, where $m_e$ is the electron mass. For $v \sim 0.1$~eV, this imitates the coupling strength of a standard hidden axion with a decay constant of order $10^{10}$~GeV. This may be interesting for future experimental searches in Shining Light Through Walls type experiments, which are currently at a sensitivity of $\sim 10^{7}$~GeV for small ALP masses and will reach a sensitivity of $\sim 10^{11}$~GeV, however only for very light ALPs with masses of $\sim 10^{-4}$ eV \cite{Redondo2010}.

Notice that the above estimate for the strength of the effective coupling is only valid for very low energy photon-graviaxion processes. In high-energy processes, for example, in the graviaxion production in stars, the coupling is expected to be much stronger suppressed because of a very low compositeness scale of the $\eta_{\nu}$ boson and the high-energy softening of the gravitational vertex. Due to this, we should not expect the standard axion-type correlation between the predictions for Shining Light Through Walls and solar axion experiments (see, e.g., \cite{Irastorza2012}).

\subsection{Short-distance gravity measurements}   
 
The connection between the scale of the neutrino Compton wavelength and the experimental frontier of gravitational force measurements was already established in the past \cite{Dvali1999}. In the considered scenario, the physics that set the neutrino mass simultaneously modified Newton's law due to extra dimensions \cite{Arkani-Hamed1998}.    

The neutrino mass mechanism of \cite{Dvali2016}, which generates the neutrino masses of the order of the scale of nonperturbative gravity $\Lambda_G$, offers another way of realizing a connection between the neutrino mass and the current experimental frontier of short-distance tests of gravity \cite{Adelberger2009}. Now it is natural to ask whether we can predict any observable corrections to Newtonian gravity at distances shorter than $\Lambda_G^{-1}$. In general, it would be hard to make a concrete prediction due to the lack of knowledge of a direct relation between the correlator \eqref{RRcorr} and the modification of the graviton propagator. However, in the Domestic Axion scenario presented above, one very concrete prediction emerges that is directly tied to the generation of the up-quark mass by the neutrino condensate. 

Indeed, we predict a new force mediated by the Higgs-like excitation(s) of the neutrino condensate, which describe(s) small fluctuations of its absolute value. For illustrating this point, it is enough to consider one of such modes, which we shall denote by $\sigma_{\nu}$. Then, the expansion of the neutrino condensate around its VEV can be written as 
\begin{equation}
\bar{\nu}_L\nu_R = (v + \sigma_{\nu}) e^{i {\eta_{\nu} / v}}.
\end{equation}
Due to the UV-softening of the gravitational vertex (\ref{vertex}), the coupling of $\sigma_{\nu}$ to a constituent up quark inside the proton is suppressed by powers of the ratio $(v/m_P)$, where $m_P$ is the proton mass. However, at the same time it is enhanced by the same parameter $\xi$ that is responsible for the generation of the relatively large up-quark mass. Hence, the effective coupling to a proton, up to an unknown coefficient of order one, is expected to be given by $\xi (v/m_P)^{\beta}$, where $\beta$ is a positive number which parameterizes the softening of the gravitational vertex in high-energy processes. 

At distances $r$ around or larger than the Compton wavelength of the $\sigma_{\nu}$ boson, its exchange will result into a gravity-like potential between two protons of the order $V(r)_{\nu} \sim  \xi^2 (v / m_P)^{2\beta} (e^{-rm_{\sigma_{\nu}}} / r)$, where $m_{\sigma_{\nu}}$ is the mass of the $\sigma_{\nu}$ boson. Since $\xi\, v \sim m_u$ and $v \sim m_{\nu}$, we can rewrite the new force in terms of the quark and neutrino masses as
\begin{equation}
V(r)_{\nu} \sim  \left({m_u \over m_{\nu}} \right )^2\left ({m_{\nu} \over m_P} \right )^{2\beta}{e^{-rm_{\sigma_{\nu}}} \over r} \, .
\label{force1}
\end{equation} 
Putting this into the conventional expression for gravity-competing forces, 
\begin{equation}
V(r)=-G_N \frac{m_1 m_2}{r}\left(1+\alpha e^{-r/\lambda}\right)\, ,
\end{equation}
we obtain for two protons, $m_1=m_2=m_P$, the parameters $\alpha\sim 10^{128 - 58\beta}$ and $\lambda=m_{\sigma_{\nu}}^{-1}$.

For $m_{\sigma_{\nu}} \sim v \sim 0.1$~eV, the existing experimental measurements \cite{Chen2016} put the bound $\beta \gtrsim 2.1$.\footnote{Notice, the case $\beta =2$, which corresponds to a simplest minimal suppression that one can obtain based on very general scaling arguments, is compatible with the current bounds and can lead to observable effects for a slightly higher mass of $\sigma_{\nu}$.}

The interesting message we would like to take from here is that one place to look for the effects of the $\sigma_{\nu}$ boson is in searches for a new force at micron and sub-micron distances, which can exceed the strength of Newtonian gravity by many orders of magnitude. The force is highly sensitive to the parameter $\beta$, which we cannot predict, but for values $\beta \simeq 2$ such a force can be just within the reach of the planned improved measurements \cite{Nesterov2013-Schmidt2016}. The modification of Newton's law must appear as a threshold effect, which should diminish both above and below the scale $r \sim m_{\sigma_{\nu}}^{-1}$. As it is clear from (\ref{force1}), for $r \gg m_{\sigma_{\nu}}^{-1}$ the force diminishes exponentially. Instead, for distances $r \ll m_{\sigma_{\nu}}^{-1}$ it is expected to diminish as a power-law, e.g., for $\beta = 2$, 
\begin{equation}
V(r)_{\nu} \sim  \left ({m_um_{\nu} \over m_P^2}  \right )^2{v^4r^3} \, ,
\label{force2}
\end{equation}
due to the decoupling of IR physics from short-distance effects.

Also note that, since the $\sigma_{\nu}$ boson has a dominant coupling to the up quark, the resulting force will depend on the number of up quarks in the source and thus is predicted to be {\it isotope-dependent}. Namely, the coupling to a proton is by a factor of two stronger than the coupling to a neutron.\footnote{Note that the force will continue to be isotope-dependent even in a nonminimal scenario in which also the down-quark mass is generated from the neutrino condensate, since the relative strength of the coupling to up and down quarks will be set by the ratio of the quark masses. Thus, in this case the coupling to the down quark will be stronger and correspondingly the coupling to a neutron will be stronger than to a proton.}

\subsection{Majorana versus Dirac?}

One may ask whether our model predicts the Dirac nature of neutrinos. Unfortunately, we cannot make such a definite prediction, since the scenario can work also for a Majorana neutrino. Indeed, even if only one active left-handed neutrino is introduced, $\nu_L$, there still exists a chiral symmetry anomalous under gravity, which acts on $\nu_L$ and thus leads to neutrino condensation. Since $\nu_L$ is a part of a lepton doublet, $L \equiv (\nu_L, e_L)$, the neutrino condensate $\langle \nu_L C\nu_L \rangle $ transforms as a triplet under the weak $SU(2)$ symmetry. Nevertheless, an effective doublet can be composed by convoluting it with a doublet quark condensate, and the up-quark mass can still be generated through the following operator:
\begin{equation}
(\bar{Q}_L^j u_R )(L_j C L_m) (\bar{Q}_L^m u_R) \, , 
\label{higherterm}
\end{equation} 
where $j,m=1,2$ are the indexes of the weak $SU(2)$ gauge symmetry and $Q \equiv (u_L,d_L)$ is the quark doublet.

\subsection{Flavor physics}\label{sec:Flavor}

In the present analysis, we did not include other charged fermions, but they can be easily incorporated by adding additional fermion legs to the effective vertex (\ref{vertex}). Such a vertex will generate an additional contribution to the masses of all the fermions once the gravity-induced fermion condensates are taken into account. Without taking extra care, the resulting corrections to the masses will naturally be of the order of the neutrino masses.  

A phenomenologically interesting possibility from the point of view of flavor physics opens up in case when the new mass contributions are not diagonal in the eigen-basis of the SM Higgs Yukawa couplings. In such a case, the new flavor-changing processes emerge from the IR neutrino sector.   

Remarkably, even if the IR flavor violation at the scale of the neutrino masses is order one, such a possibility can nevertheless be fully viable phenomenologically and potentially testable. This may come as a surprise, since generating masses from sources other than a single Higgs condensate is normally associated with severe problems, such as flavor-changing neutral currents. In our case, the role of the second Higgs doublet with a tiny VEV is played by the neutrino condensate. The reason why this condensate {\it a priori} is not causing the usual problems, such as the flavor-changing neutral currents mediated by the exchange of the $\sigma_{\nu}$ boson, is because its compositeness scale is extremely low: even if the $\sigma_{\nu}$ boson has large (i.e., order one) flavor-nondiagonal couplings, it decouples very efficiently from the high-energy processes. Correspondingly, the contribution of the neutrino composites into the high-energy flavor-changing processes, such as e.g., $K^0-\bar{K}^0$ transitions or $\mu\rightarrow e  +\gamma$ decays, is small, but can be potentially interesting for future measurements, as examined in more detail in App. \ref{App:Flavor}.

The above also raises the question whether we can generate the masses of other light charged fermions entirely via the mechanism considered here, as an alternative of generating their masses from the coupling to the Higgs doublet. However, since all the effective masses generated through the neutrino condensate are not present in the early Universe before CMB formation \cite{Dvali2016}, our effective mass generation mechanism can only account for the entire masses of neutrinos, up and down quarks, while all other fermions need to have additional mass sources.

\subsection{Topological defects} 
 
It is well known that a phase transition with PQ symmetry breaking can form axionic cosmic strings. These strings later become boundaries of domain walls \cite{Vilenkin1982} and decay producing axions. In our case, everything happens at the same scale: the generation of the quark condensate and the generation of the $\eta'$ mass take place around QCD temperatures. So the axionic cosmic strings are produced in form of small loops spanned by membranes (domain walls) and decay very quickly. The second phase transition around the neutrino mass scale also forms cosmic strings bounded by walls. This time, the string-wall system is composed out of the graviaxion $\eta_{\nu}$, but since $\eta_{\nu}$ can in principle be substantially lighter than the VEV of the neutrino condensate, these strings can be parametrically longer-lived before they decay into $\eta_{\nu}$-s and neutrinos. Just like in the case of standard axionic strings, this could be a way of populating the Universe by a large number of $\eta_{\nu}$ particles. The precise $\eta_\nu$ density and thus its contribution to the dark matter abundance in our Universe strongly depends on the free parameters of our model, such as the exact value of the scale $\Lambda_G$.

\section{Conclusions} 

It is a standard lore that new physical effects can hide and decouple if the energy scale of their origin is very high. In particular, the standard hidden axion is decoupled because of an extremely high scale of PQ symmetry breaking. Apart from the new naturalness problem in form of the hierarchy between the PQ and the weak scales, this leaves us with the questions why a whole new high-energy sector should be designed solely with the purpose of nullifying a particular parameter of the SM?  

In the present note, we have proposed an alternative hiding place for axion physics within the SM in form of a deep-infrared scale, without the need of postulating any new particle species. This IR scale is related to the neutrino masses. Our axion consists of the $\eta'$ meson with a minuscule admixture of the neutrino composite $\eta_{\nu}$. The latter is a pseudo-Goldstone of the neutrino condensate triggered by nonperturbative gravity and gets its mass from the gravitational anomaly. 

The neutrino condensate does several jobs. On the one hand, it generates the mass for the up quark {\it spontaneously}. This is the key that in our scenario allows the $\eta'$ meson to act as an axion and cancel the $\theta$-term. On the other hand, the Goldstone boson $\eta_{\nu}$ originating from the neutrino condensate protects the shift symmetry of the $\eta'$-axion from being broken by the gravitational anomaly \cite{Dvali2013}. At the same time, the neutrino condensate is a natural source for generating the neutrino masses via the scenario proposed in \cite{Dvali2016}.

However, the latter possibility is not tied to the Domestic Axion scenario presented here, for which it is enough that only a single neutrino flavor gets its mass from the gravitational anomaly, whereas the other flavor masses can be generated in conventional ways. In such a case, the field content of the model is reduced to \cite{Dvali2013}, in which the bare mass of a single neutrino is set to zero. In this situation, the Domestic Axion scenario would be fully realized, but the possibility of explaining the masses of all the neutrinos from the gravitational mechanism would not be used.

Conversely, the neutrino condensation and the neutrino mass generation model presented in \cite{Dvali2016} can be used without the need of much fine-tuning in the neutrino sector, even if one is not willing to address the strong $CP$ problem. The introduction of the parameter $\xi\sim 10^7$ is only needed if we want to spontaneously generate the up-quark mass by the neutrino condensate and thus solve the strong $CP$ problem by the Domestic Axion scenario described in the present paper. As explained in the text, this is not increasing the number of required tunings:  we trade the tuning of the Yukawa coupling constant for the tuning of $\xi$, but with a big bonus of solving the strong $CP$ problem.  

In this light, it is natural as well as beneficiary to unify the two scenarios that nicely complete each other and connect the solution to the strong $CP$ problem and the origin of the neutrino masses to a single gravitational source. \\

From the broader perspective, what we have observed is that a very low scale compositeness can mask new physical effects not less (and in some cases even more) efficiently than the phenomenon of high-energy decoupling. This is a very general message that we believe should be payed more attention to when looking for new physical effects.

\section*{Acknowledgements}

We thank Cesar Gomez and Georg Raffelt for valuable discussions. We also thank Helena Schmidt for discussions on short-distance gravity measurements. The work of G.~D. was supported in part by the Humboldt Foundation under Humboldt Professorship, ERC Advanced Grant 339169 "Selfcompletion", by TR 33 "The Dark Universe", and by the DFG cluster of excellence "Origin and Structure of the Universe". The work of L.~F. was supported by the International Max Planck Research School on Elementary Particle Physics.

\appendix

\section{Axion and graviaxion mass matrix}\label{App:MassMatrix}

In this appendix, we shall explicitly show how the $\eta'$ and the $\eta_{\nu}$ cancel both the QCD and the gravitational $\theta$-terms and shall diagonalize their mass matrix. We will achieve this by using the three-form formalism \cite{Dvali2005_2}. A detailed discussion of the diagonalization of the mass matrix in case of mixing the $\eta_{\nu}$ meson with a conventional axion is given in \cite{Dvali2013}. The only difference in our case is that the standard axion is replaced by $\eta'$. 

The three-form formalism uses the fact that a nonzero topological vacuum susceptibility both in gravity and in gauge theory implies that the topological density can be interpreted as the gauge-invariant field strength of a massless three-form field: $R\tilde{R} \equiv dC_G \equiv E_G$ and $G\tilde{G} \equiv dC \equiv E$, where $C\equiv AdA - {2 \over 3} AAA$ and $C_G \equiv \Gamma d\Gamma - {2 \over 3} \Gamma\Gamma\Gamma$ are the QCD and gravitational Chern-Simons three-forms, respectively. 

The low-energy effective theory that fully captures the details of the mass-gap generation is a gauge invariant theory of these three-forms coupled to pseudo-Goldstone bosons of anomalous currents. The gauge invariance and anomaly fully determines the form of this effective Lagrangian.
In the present case we have
\begin{eqnarray}\label{mixing1}
 \mathcal{L} &=& \frac{1}{2V^4}E^2 +\frac{1}{2v^4} E_G^2 \, - \, \frac{\eta'}{V}E- \left(\frac{\eta' }{V} + \frac{\eta_\nu}{v} \right ) E_G \nonumber\\
&&+ \, \frac{1}{2}\,\p_\mu \eta'\p^\mu\eta'\,  + \frac{1}{2}\, \p_\mu \eta_\nu\p^\mu\eta_\nu \, .
\end{eqnarray}
As shown in \cite{Dvali2005_2}, the higher-order polynomial terms in $E$ and $E_G$ can easily be taken into account and they only affect the form of the resulting pseudo-Goldstone potentials for large field values, but cannot affect the mass gap. The terms with higher derivatives vanish for constant field values (i.e., in the zero momentum limit) and cannot affect the form of the scalar potentials. Hence, they are irrelevant (see \cite{Dvali2005_2} for details). We do not explicitly display the numerical coefficients that absorb irrelevant combinatoric factors, especially in the light of the strong hierarchy between the scales, $V \gg v$.

The equations of motion for $C$ and $C_G$ are 
\begin{eqnarray}\label{equE}
\mathrm{d}\left( E - V^3\eta'\right)&=&0\, ,\nonumber\\
\mathrm{d}\left( E_G -  v^4\left ( \frac{\eta'}{V} + \frac{\eta_\nu}{v}\right ) \right)&=&0\, ,
\end{eqnarray}
and the ones for $\eta'$ and $\eta_\nu$ read 
\begin{eqnarray}\label{eqscalars}
\Box \eta' + {1 \over V} ( E_G  + E) &=& 0\, , \nonumber\\
\Box\eta_\nu + \frac{1}{v} E_G &=& 0\, .
\end{eqnarray}
Integrating (\ref{equE}) we get
\begin{eqnarray}
E &=&  V^4\left (\frac{\eta'}{V}  + \theta\right) , \nonumber\\
E_G &=& v^4 \left ( \frac{\eta'}{V} + \frac{\eta_\nu}{v}  \, + \theta_G \right ) , 
\label{electric} 
\end{eqnarray} 
where $\theta$ and $\theta_G$ appear as two arbitrary integration constants. Notice, in the absence of the $\eta_{\nu}$ boson, there would be no way to compensate both $\theta$-terms by a shift of $\eta'$ alone. This is a simple manifestation of how gravity ruins the solution to the strong $CP$ problem by ``destructing" the axion (in the present version $\eta'$) from its job of compensating the $\theta$-angle of QCD. 

However, as we can easily see, the problem is solved by $\eta_{\nu}$. Namely, both integration constants $\theta$ and $\theta_G$ can be eliminated by the appropriate shifts of $\eta'$ and $\eta_{\nu}$:  $\eta' \rightarrow \eta'  -  V \theta, ~~ \eta_{\nu}  \rightarrow \eta_{\nu} -   v(\theta - \theta_G)$. Moreover, the vacuum of the theory is at  
\begin{equation}
\eta' = - V \theta \, , ~~~ \eta_{\nu} =   v(\theta - \theta_G)\,, 
\label{vacuum}
\end{equation} 
where $E=E_G=0$ and both topological susceptibilities vanish. The physical meaning of this is that both three-forms $C$ and $C_G$ become massive by eating up the corresponding pseudo-Goldstone bosons, $\eta'$ and $\eta_{\nu}$. 

After eliminating the two integration constants, we can plug the expressions (\ref{electric}) for $E$ and $E_G$ into (\ref{eqscalars}) and get the following effective mass terms: 
 \begin{eqnarray}\label{equmasses}
\Box \eta' + V^2 ( 1  +  \epsilon^4 ) \eta'  +  \epsilon v^2 \eta_\nu  &=& 0 \, ,\nonumber\\
\Box\eta_\nu + \epsilon  v^2 \eta'  + v^2 \eta_\nu   &=& 0\, .
\end{eqnarray} 
Ignoring terms of order $\epsilon^4$, the corresponding mass terms in the Lagrangian are
\begin{equation}
\mathcal{L}_{\rm mass} \, = \, - {1 \over 2} V^2 \eta^{'2} \,  - \,  \epsilon v^2 \eta'\eta_{\nu}  \, -   
\,  {1 \over 2} v^2 \eta_{\nu}^{2} \, .
\label{masses}
\end{equation} 
As one can deduce from the mass matrix
\begin{equation}
M^2 = V^2
\begin{pmatrix}
1 & \epsilon^3 \\
\epsilon^3 & \epsilon^2
\end{pmatrix}
\end{equation}
(where we again omitted an irrelevant combinatoric factor), the mixing between the two states is absolutely minuscule ($\sim \epsilon^3$). Therefore, the eigenvalues of the mass matrix are approximately given by
\begin{equation}
m^2_{1,2}\simeq \frac{1}{2}(V^2+v^2)\pm\frac{1}{2}(V^2-v^2)\, ,
\end{equation}
and the corresponding eigenstates (up to $\sim \epsilon^3$ mixing) are $\eta'$ and $\eta_{\nu}$, 
\begin{equation}
a_{\rm PQ} \, = \, \eta'  + {\mathcal O}(\epsilon^3) \eta_{\nu} \,, ~~~ 
a_{G} \, = \, \eta_{\nu}\,  + \, {\mathcal O}(\epsilon^3) \eta'\, ,
\label{states2}
\end{equation}
with masses $m_{\eta'}^2 = V^2$ and $m_{\eta_\nu}^2 = v^2$, respectively.

\section{Structure of gravitationally induced fermion condensate}\label{App:Condensate}

Let us first ignore all the SM gauge and Higgs interactions and consider gravity coupled to a certain number $N_F$ of fermion flavors, $\psi_i$ and $\psi_{c\bar{i}}$ with $i,\bar{i}=1,2,...\,,N_F$, where we wrote all the fermions in the left-handed basis and the subscript $c$ stands for anti-fermion. For example, in the massless limit of the SM with three right-handed neutrinos included, we have $N_F = 24$. 

From the anomaly and topological arguments \cite{Dvali2005,Dvali2013} discussed in the text, we know that the fermions must condense and spontaneously break the anomalous chiral symmetry
\begin{equation} 
\psi_i \rightarrow e^{i \alpha} \psi_i \, ,  ~~  \psi_{c\bar{i}}  \rightarrow e^{i \alpha} \psi_{c\bar{i}}  \, .
\label{gravichiralNF} 
\end{equation}  
However, we do not have any definite information about the flavor structure of the condensate. This structure must be determined dynamically by minimization of the effective potential for the following order parameters: 
\begin{equation}
\hat{X}_{i\bar{j}} \equiv \psi_i\psi_{c\bar{j}}\, , ~ X_{ij} \equiv \psi_i C\psi_{j}\, , ~\bar{X}_{\bar{i}\bar{j}} \equiv \psi_{c\bar{i}}C\psi_{c\bar{j}}\, ,
\label{ABC}
\end{equation} 
where $C$ is the matrix of charge conjugation. Notice, although the fermions $\psi_i, \psi_{ci}$ can be embedded as a fundamental representation of the $U(2N_F)$ group, the Lorentz-invariant bilinear order parameters form the representations of the $U(N_F)_L\times U(N_F)_R$ group acting on indexes $i$ and $\bar{i}$, respectively: $\hat{X}_{i\bar{j}}$ is bifundamental, whereas $X_{ij}$ and $\bar{X}_{\bar{i}\bar{j}}$ transform as symmetric tensors under $U(N_F)_L$ and $U(N_F)_R$, respectively.  

We can classify various terms in the effective potential according to their transformation properties with respect to the $U(N_F)_L\times U(N_F)_R$ flavor group. Namely, we split all possible terms in two categories: the terms that are flavor invariants and the terms that break one part or an entire flavor group explicitly. 
   
There is a finite number of independent invariants, which have the form of various traces, such as, $\Tr (\hat{X}^+\hat{X})$, $\Tr (\hat{X}^+\hat{X} \hat{X}^+\hat{X})$, $...\,$, $\Tr(X^+X)$, $\Tr(\bar{X}^+\bar{X})$,  
$\Tr(\hat{X}X^+\bar{X}\hat{X}),$ $...\,$. The effective potential can in general represent an infinite polynomial of such invariants scaled by powers of $\Lambda_G$. 

In order to characterize the terms that break the flavor group explicitly, we need some guideline. As such, we are going to use the anomaly. It is reasonable to expect that pure gravitational effects only explicitly break the anomalous chiral symmetry $U(1)_{G}$, and leave invariant the anomaly-free subgroup $Z_{2N_F}$ as well as the discrete symmetry under the exchange of fermions and anti-fermions. An operator with such transformation properties is
\begin{equation}
\epsilon^{i_1...\,i_{N_F}} \epsilon^{\bar{j}_1...\,\bar{j}_{N_F}}  
\hat{X}_{i_1\bar{j}_1}...\,\hat{X}_{i_{N_F}\bar{j}_{N_F}}\,, 
\label{breaker}
\end{equation}
which is analogous to the 't Hooft vertex in QCD. Of course this operator will in general be multiplied by an arbitrary function $f$ of the phase-independent invariants.

Again, we must stress that we are {\it not} making any assumption about the possible origin of the above vertex from gravitational instantons. The analogy with the instanton-induced 't Hooft vertex in QCD is purely from the point of view of its symmetry properties: if the gravitational anomaly generates a mass gap for the $\eta_{\nu}$ pseudo-Goldstone, the effective  potential of the order parameters must contain terms that break the $U(1)_G$ symmetry explicitly down to $Z_{2N_F}$. This uniquely fixes the structure of the minimal vertex (\ref{breaker}), irrespective of its underlying origin, which can be fully quantum rather than semi-classical.      

After including all possible terms, we get an effective potential invariant under $SU(N_F)\times SU(N_F)\times U(1)_{V}\times Z_{2N_F}$ symmetry. The form of the condensate that spontaneously breaks this symmetry group is determined by minimization of the potential. It is well accepted that the analogous potential in case of QCD  breaks the flavor group down to a diagonal subgroup $U(N_F)_{V}$. However, {\it a priori} there is no reason that gravity should follow the same pattern of symmetry breaking. In fact (as also discussed in \cite{Dvali2016}), it is easy to see that already an effective potential that includes up to quartic order invariants in the order parameters $\hat{X}$, $X$, and $\bar{X}$ allows for a very rich variety of patters of flavor symmetry breaking. The possibility of spontaneous breaking of the flavor group is very important due to resulting new flavor-violating phenomena. This will be analyzed in more details in the next appendix. 

Let us now switch on the SM gauge and Higgs interactions. These break the flavor group explicitly down to a much smaller subgroup. In particular, after ``dressing" the effective gravitational vertex (\ref{breaker}) by effects of QCD and electroweak interactions, we can integrate out all the heavy species of masses $\gg \Lambda_G$ and obtain an effective vertex for the species that are getting masses from the gravitational effects, i.e., the up quark and the neutrinos. This resulting effective vertex has the form (\ref{vertex}) and is enough for reducing our solution to the strong $CP$ problem to its bare essentials. However, for precision phenomenology, taking into account other species is important as discussed in the following appendix.

\section{Flavor-violating processes}\label{App:Flavor}

Let us estimate flavor-violating neutral currents due to our modified neutrino sector. From the point of view of such flavor violation, the story effectively reduces to the introduction of additional Higgs doublets, which are the composites of the left-handed lepton doublets $L^{\alpha} \equiv (\nu^{\alpha}_L , e^{\alpha}_L)$ and the hight-handed neutrinos $\nu_R^{\alpha}$, where $\alpha,\beta = 1,2,3$ are generation (family) indexes.
  
In order to understand the essence of flavor violation, it is enough to consider only one such doublet,  $\equiv(\bar{L}^{\alpha} \nu_R^{\beta})$. Notice that the effective doublet is not necessarily diagonal in family space.  

The neutral component of this composite doublet is the neutrino bilinear, which develops a VEV. Expanding around its VEV, we can write $\bar{\nu}_L\nu_R = (v + \sigma_{\nu}) e^{i {\eta_{\nu} / v}}$, where $\sigma_{\nu}$ describes excitations of the absolute value and plays the role analogous to the neutral Higgs particle, $h_0$. If the couplings of $h_0$ and $\sigma_{\nu}$ to quarks of the same charge are not diagonal in the mass-eigenstate basis, there will be flavor-changing neutral currents mediated due to their exchange. 

In order to trace the origin of flavor violation more explicitly, let us first consider a minimal coupling needed for accomplishing our axion mechanism and see that it is not leading to flavor violation. In this minimal scheme, it is enough to consider the case in which only one neutrino transforms under the $U(1)_{\rm PQ}$ symmetry, for instance, the $\nu_{\tau}$ neutrino.   

Consider a gravity-generated coupling,
\begin{align}  \label{minimal}
{ 1 \over \Lambda_G^2} (\bar{u}_R Q_L)(\bar{L}^3\nu_{\tau R} ) \, = \,  &{ 1\over \Lambda_G^2}  (\bar{u}_Ru_L)(\bar{\nu}_{\tau L}\nu_{\tau R} )\,  \nonumber\\
&+{ 1 \over \Lambda_G^2}  (\bar{u}_Rd_L)(\bar{\tau}_L\nu_{\tau R}) \, ,  
\end{align}  
where $Q \equiv (u_L,d_L)$ is the first-generation left-handed quark doublet. For the purpose of the discussion of flavor conservation, the function $f$ of the invariants introduced in (\ref{vertex}) is not important and we drop it for simplicity. The anomalous PQ symmetry in this case can be identified as the chiral symmetry acting on $u_R$ and $\nu_R$ species only, 
\begin{equation} 
u_R \rightarrow e^{i \alpha} u_R \, ,  ~~  \nu_R \rightarrow e^{i \alpha} \nu_R \, .
\label{gravichiralR} 
\end{equation}  
This symmetry is incompatible with the Yukawa couplings of $u_R$ and $\nu_R$ to the Higgs doublet $H$. Correspondingly, unlike the rest of the fermions, the up quark and the $\nu_\tau$ neutrino are not getting any mass from the VEV of the Higgs. 

In such a case, the couplings of both the neutral Higgs $h_0$ as well as of the $\sigma_{\nu}$ are diagonal in the mass-eigenstate basis and no flavor-violating neutral currents appear. Notice, the last term in (\ref{minimal}) can contribute into the decay of the $\tau$ lepton into a pion and a neutrino, but since the vertex is strongly suppressed at high energies, the rate is expected to be negligible. For instance, already for a suppression by a factor of $ v^2 / m_{\tau}^2$, the rate is way beyond current experimental sensitivity. This suppression of an effective vertex in high-energy processes is the main reason for making this new IR physics compatible with present experimental bounds.\\     
 
Let us now turn to a generic nonminimal case, in which all three generations can be involved in the PQ symmetry breaking as well as in the efffective gravitational vertex.  

Before illustrating in details, let us summarize the story. As mentioned above, the flavor-violating neutral currents will appear if the Yukawa coupling matrixes of the $\sigma_{\nu}$-s and the $h_0$ are not diagonal in the fermion mass eigenbasis. In such a case, it is useful to split the potential flavor-violating contributions into the ones mediated by the SM neutral Higgs $h_0$ and the ones mediated by the $\sigma_{\nu}$ bosons. These are both suppressed, but because of different reasons: 
  
The Higgs-mediated flavor violation is typically suppressed by a factor of $\delta m_{\alpha\beta}^2  / |m_{\alpha} - m_{\beta}|^2$, where $m_{\alpha}$ are the fermion mass eigenvalues coming from the Higgs Yukawa couplings and $\delta m_{\alpha\beta}$ is the off-diagonal mass generated by the neutrino condensate. The flavor violation mediated by $\sigma_{\nu}$ is universally suppressed due to the suppression of the effective gravitational vertex in high-energy processes. We assume here that this suppression goes as powers of $v^2 / E^2$, although it could in principle be stronger.   

All the above can only take place if gravity violates flavor, that is, if gravity generates off-diagonal effective couplings for the $\sigma_{\nu}$-s in the basis in which Higgs Yukawa couplings are diagonal. {\it A priori} we have no way of predicting this. However, we can make some useful parameterization of our ignorance. 
 
The generation of off-diagonal couplings by gravity can be explicit or spontaneous. Since the fermion flavor group is not anomalous with respect to gravity, the explicit breaking must come from some other quantum gravity effects, which we can only parameterize. 

Spontaneous breaking is simpler to visualize. For spontaneous generation it is necessary that the condensates of charged leptons and quarks are off-diagonal in the basis in which the Higgs Yukawa couplings are diagonal. This depends on the minimization of the effective potential for these order parameters, and it is easy to come up with prototype potentials that would result into disoriented condensates in flavor space.  
  
Let us estimate the flavor violation in an example of the down-quark sector. The effective Yukawa coupling matrixes are
\begin{equation}
(V_h +  h_0) g_{\alpha\beta} \bar{d}_L^{\alpha}d_R^{\beta}   \, + 
(v+\sigma_{\nu}) g^{\sigma}_{\alpha\beta} \bar{d}_L^{\alpha}d_R^{\beta}\;,
\label{flavorV} 
\end{equation} 
where $\alpha,\beta = 1,2,3$ are flavor indexes and $V_h \sim 100$~GeV is the Higgs VEV. Let us work in the basis in which the SM Higgs Yukawa coupling matrix $g_{\alpha\beta}$ is diagonal. Then, if the down-quark condensate can have off-diagonal values in this basis, $g^{\sigma}_{\alpha\beta}$ will develop off-diagonal elements. The standard QCD condensate of quarks is diagonal in the mass-eigenstate basis, so the off-diagonal contribution must come from gravity. We do not know how strong such a contribution is, so we can parameterize it as unknown. 

Let us consider the $1-2$ transition via the condensate $\langle \bar{d}_L^{1}d_R^2 \rangle \equiv \langle \bar{d}_Ls_R \rangle$. We shall assume that the condensate as well as the Yukawa matrixes are $L-R$ symmetric. The resulting off-diagonal Yukawa coupling of $\sigma_{\nu}$ is $g^{\sigma}_{12} \sim  {v^{-3}\langle \bar{d}_Ls_R \rangle} $ and this induces a shift in the off-diagonal mass, $\delta m_{12} \sim {v^{-2}\langle \bar{d}_Ls_R \rangle}$. This generates the flavor-changing neutral currents via exchanges of $h_0$ and $\sigma_{\nu}$. 

Let us compute the first one. This will be controlled by the effective off-diagonal coupling to the quarks that $h_0$ will acquire after we re-diagonalize the small off-diagonal mass term, $\delta m_{12}$, induced by the neutrino condensate. The new mixing angle is suppressed by the ratio of the this off-diagonal mass to the diagonal mass difference, $\delta m_{12}/(m_s -m_d)\simeq \delta m_{12}/m_s$, so that we obtain $g_{sd} \sim (m_s / V_h) (\langle \bar{d}_Ls_R \rangle / (m_s v^2))$. Thus, the $(\bar{s}d)^2$-operator induced by the Higgs exchange has the form
\begin{equation}
(\bar{s}d)^2 { 1\over m_h^2} \left({\langle \bar{d}_Ls_R \rangle \over V_h v^2} \right )^2 ,  
\label{flavorHiggs1} 
\end{equation} 
where $m_h$ is the Higgs mass. Even if we assume that the off-diagonal condensate is of the same order as the diagonal one, the condensate must be suppressed by the masses of the quarks relative to the scale $v$. For example, even if we assume 
\begin{equation}
\langle \bar{d}_Ls_R \rangle \sim v^3{v \over \sqrt{m_sm_d}}\,,
\label{offdiag}
\end{equation}
the operator (\ref{flavorHiggs1}) is hugely suppressed.  

The similar operator generated by the exchange of a $\sigma_{\nu}$ boson will have the form
\begin{equation}
(\bar{s}d)^2 { 1 \over m_{\sigma}^2} \left({\langle \bar{d}_Ls_R \rangle \over v^3} \right )^2 \sim (\bar{s}d)^2 { 1 \over m_{\sigma}^2} {v^2 \over m_sm_d} \, .
\label{flavorHiggs2} 
\end{equation} 
Since $m_{\sigma} \sim v$, this operator looks pretty strong, but of course we have to remember that this is an effective interaction valid only at energies below the neutrino mass scale $v$. So the contribution into high-energy processes, such as $K^0 - \bar{K}^0$ transitions, is additionally suppressed by the ratio of the scales $v^2 / m_K^2$, which gives another factor of order $10^{-20}$. Overall, we are down to an effective scale of $(\bar{s}d)^2 / (10^{16}~{\rm GeV^2})$, which although suppressed is stronger than the previous one and can be of some phenomenological interest.\\

Analogously, we can estimate the processes with lepton-flavor violation. Consider a leptonic fragment of the gravitational vertex that involves charged leptons of the first two generations and neutrinos of the third generation (with all other fermion pairs being replaced by their masses and VEVs),   
\begin{equation}
{ 1 \over \Lambda_G^5} (\bar{e}_L\mu_R)(\bar{\mu}_L e_R) (\bar{\nu}_{\tau L} \nu_{\tau R}) \, . 
\label{leptons} 
\end{equation}     
We assume this is written in the basis in which the Higgs Yukawa couplings to the charged leptons are diagonal. If in this basis the condensate $\langle \bar{\mu}_Le_R \rangle$ is nonzero, this results into the following strength of the off-diagonal couplings of the Higgs boson and $\sigma_{\nu}$ with charged leptons:
\begin{equation} 
\label{Asymmlepton}
 h_0 \bar{\mu}_L e_R \left ({\langle \bar{\mu}_Le_R \rangle \over V_hv^2} \right) +  
\sigma_{\nu} \bar{\mu}_L e_R \left ({\langle \bar{\mu}_Le_R \rangle \over v^3} \right).
\end{equation}
The first coupling at one loop can result into $\mu \rightarrow e + \gamma$ decay, whereas the second one into a direct decay of a muon into an electron and a neutrino-antineutrino pair. Again, we have to take into account the additional suppression by a factor of $v^2 / m^2_{\mu}$, due to the decoupling of IR physics in high-energy processes. This decoupling is the key of putting even a maximal IR flavor violation into a potentially phenomenologically interesting  domain.

\end{document}